\documentclass[twocolumn,english,conference]{IEEEtran}
\ifCLASSINFOpdf
\else
\fi

\usepackage[T1]{fontenc}
\usepackage{babel}
\usepackage{array}
\usepackage{calc}
\usepackage{amsmath}
\usepackage{amssymb}
\usepackage{amsthm}
\usepackage{graphicx}
\usepackage{cite}
\usepackage{subfigure}
\usepackage{color}

\usepackage{enumitem}
\usepackage{multirow}
\usepackage{hhline}
\usepackage[ruled]{algorithm2e}
\usepackage{algpseudocode}

\usepackage[mathlines,switch]{lineno}

\makeatletter

\let\oLRDforeign@language\foreign@language
\DeclareRobustCommand{\foreign@language}[1]{%
	\lowercase{\oLRDforeign@language{#1}}}

\makeatother
\allowdisplaybreaks

\makeatletter
\newcommand{\thickhline}{%
	\noalign {\ifnum 0=`}\fi \hrule height 1.5pt
	\futurelet \reserved@a \@xhline
}
\newcolumntype{"}{@{\hskip\tabcolsep\vrule width 1pt\hskip\tabcolsep}}
\makeatother

\newcommand\blfootnote[1]{%
	\begingroup
	\renewcommand\thefootnote{}\footnote{#1}%
	\addtocounter{footnote}{-1}%
	\endgroup
}

\begin{document}
	%
	\title{Can Predictive Filters Detect Gradually Ramping False Data Injection Attacks Against PMUs?}

	
	
	%
	\author{\IEEEauthorblockN{
			Zhigang Chu,
			Andrea Pinceti,
			Reetam Sen Biswas,
			Oliver Kosut,
			Anamitra Pal,
			and Lalitha Sankar}
		\IEEEauthorblockA{School of Electrical, Computer and Energy Engineering\\
			Arizona State University}}


	\maketitle
	\pagestyle{plain}
	\begin{abstract}
		Intelligently designed false data injection (FDI) attacks have been shown to be able to bypass the $\chi^2$-test based bad data detector (BDD), resulting in physical consequences (such as line overloads) in the power system. In this paper, it is shown that if an attack is suddenly injected into the system, a predictive filter with sufficient accuracy is able to detect it. However, an attacker can gradually increase the magnitude of the attack to avoid detection, and still cause damage to the system.
	\end{abstract}
	

	%
	\IEEEpeerreviewmaketitle
	\global\long\def\figurename{Fig.}
	\global\long\def\tablename{TABLE}

	\section{Introduction}\label{sec:Introduction}
    In the past decade, phasor measurement units (PMUs) have been widely deployed in power systems for monitoring, protection, and control purposes. Since PMUs can directly measure the bus voltage phasor with high sampling rate and accuracy, they have the potential to play a significant role in real-time power system state estimation (SE)\cite{Zhao16} 
    dynamic security assessment \cite{Zhang19, DSA_PMU2012}, system protection \cite{Neyestanaki15},
    and system awareness \cite{PMU_monitoring2010}. 
    \blfootnote{The first three authors are students ordered by contribution. The last three authors are faculty members ordered alphabetically.}

	Meanwhile, cyber-attacks against the communication and computing infrastructure of the monitoring and control systems of electric power systems have become a growing concern  \cite{StuxNet2014,UkraineAttack,USAttack2018_2}. As an increasingly important component of this infrastructure, PMUs are also prone to cyber-attacks \cite{Lin_PMUcybersecurity2012,Beasley2014_PMU}. Therefore, it is of great importance to evaluate the vulnerability of PMUs against potential cyber-attacks as well as to develop preemptive countermeasures.
	
	Here, we focus on a broad class of attacks known as false data injection (FDI), wherein an intelligent attacker replaces a subset of measurements with counterfeits. This can be accomplished against PMUs by, for example, spoofing the global positioning system (GPS) signal so as to manipulate a PMU's timestamps \cite{Gyorgy2018, Garcia2013}.
	In this paper, we do not limit ourselves to GPS spoofing attacks, but consider FDI attacks accomplished by any means.
	
	The main goals of this paper is to evaluate the effectiveness of countermeasures that use finite impulse response (FIR) predictive filters against sophisticated FDI attacks that are unobservable to current-generation bad data detectors (BDDs). In particular, our contributions are as follows:
	\begin{enumerate}
	    \item We create test FDI attacks using a bilevel optimization approach. These attacks are \emph{unobservable} in the sense that they are provably invisible to single-shot BDDs.
	    \item In order to test these attacks, we generate synthetic PMU data by isolating archetypal data profiles from real data. This technique allows us to create synthetic data to be tested in the context of the IEEE 118-bus system.
	    \item We investigate whether these attacks can be detected by taking into account temporal correlations. In particular, we use a predictive filtering approach based on the three sample quadratic prediction algorithm (TSQPA) \cite{Gao2012}, a technique that accurately predicts the next sample of real data based on the previous three. Thus, a large residue of this predictive filter indicates an attack. Moreover, we test an alternative predictive filter learned from real data, that predicts the next sample as a linear combination of the previous five.
	    \item Finally, we consider two variants of the attack, depending on whether it is applied \emph{suddenly}---i.e., at a single instant---or \emph{ramped}---i.e., gradually increased over a period of time. We demonstrate that predictive filters such as TSQPA can detect sudden attacks with high accuracy, but not ramping attacks.
	\end{enumerate}

	\section{Preliminaries}\label{sec:background}
	\subsection{PMU-based Linear State Estimation}\label{sec:LSE}
	Throughout our analysis, we assume that the power system is completely observable by PMUs. A PMU placed at a bus measures the complex voltage of that bus, and complex currents on all branches connected to it, typically at a rate of 30 samples per second \cite{Phadke2009_PMUSE}. These measurements are linear functions of the states, \textit{i.e.,} the complex bus voltages. Let $p$ be the number of buses (states), and $n$ be the number of PMU measurements in the power system, the PMU measurement vector at each time instant, $i$, is given by
	\begin{equation}
	w_i=Hx_i+e_i=\begin{bmatrix}
		I'\\ Y
	\end{bmatrix}x_i+e_i,\label{eq:ACMeasurement}
	\end{equation}
	where $w_i$ is the $n \times 1$ measurement vector; $x_i$ is the $p \times 1$ vector of true states (complex voltages); $e_i$ is an $n\times1$ additive Gaussian noise vector whose covariance matrix $R=\text{diag}[\sigma_1^2, \sigma_2^2,\ldots,\sigma_n^2]$; $H$ is the $n\times p$ measurement Jacobian matrix, consisting of $I'$, the reduced identity matrix with only rows corresponding to PMU buses; and $Y$, the dependency matrix between available current measurements and states. The weighted least squares estimate of $x_i$, $\hat{x}_i$, is given by\cite{AburBook}
	\begin{equation}
	\hat{x}_i=(H^TR^{-1}H)^{-1}H^TR^{-1}w_i.\label{eq:LinearSE}
	\end{equation}
	The conventional residue-based BDD performs $\chi^2$ test on the residue vector 
	\begin{equation}
		r_{i,S}=w_i-H\hat{x}_i \label{eq:ConvBDD}
	\end{equation}
	to detect bad measurements. Note that the subscript $S$ denotes state estimation; we introduce this notation in an effort to distinguish measurement residue resulting from state estimation from those resulting from using predictive algorithms that we introduce in Sec.~\ref{sec:TSQPA}.
	
	\subsection{Unobservable FDI Attacks on PMU Measurements}\label{sec:uFDI} 
	Suppose an attacker can change measurements in a set $\mathcal{S}$ by controlling a subset of PMUs. At each time instant, $i$, it can replace $w_i$ with
	\begin{equation}
		\bar{w}_i=w_i+d_i, \label{eq:FalseMeas}
	\end{equation} where the non-zero entries of the measurement attack vector $d_i$ are all within $\mathcal{S}$. An attack is defined to be unobservable \cite{Liu2009} to the conventional residue-based BDD if 
	\begin{equation}
		d_i=Hc_i,\label{eq:stateAtkVec}
	\end{equation}
	where the $c_i$ is the state attack vector. Substituting \eqref{eq:FalseMeas} and \eqref{eq:stateAtkVec} into \eqref{eq:LinearSE} yields the estimated states $\bar{x}_i$ under attack
	\begin{equation}
		\bar{x}_i = \hat{x}_i+c_i.\label{eq:xbar}
	\end{equation}
	The residue vector under attack
	\begin{flalign}
		\bar{r}_{i,S}&=\bar{w}_i-H\bar{x}_i\notag\\
		&=w_i+d_i-H\hat{x}_i-Hc_i=w_i-H\hat{x}_i
	\end{flalign}
	is the same as that without attack. Therefore, attacks in the form of \eqref{eq:stateAtkVec} cannot be detected by the conventional BDD.
	
	\subsection{Three Sample-based Quadratic Prediction Algorithm (TSQPA)}\label{sec:TSQPA}
	The residue-based BDD discussed in Sec. \ref{sec:uFDI} does not consider temporal correlations in PMU data to detect an anomaly. To validate the quality of the incoming measurements, Gao \textit{et al.} in \cite{Gao2012} investigate temporal correlations in PMU data to find the relationship between the past, present, and future measurements. In particular, they prove that for loads changing at a constant power factor, the complex voltage phasor follows a quadratic trajectory. Applying auto-regressive modeling on a quadratic trajectory, they show that the vector of complex voltages at the next time instant can be predicted using the present and past states as follows:
	\begin{equation}
	    x_{(i|i-1)}=3x_{i-1}-3x_{i-2}+x_{i-3}, \label{eq:TSQPA}
	\end{equation}
	where $x_{(i|i-1)}$ denotes the predicted value of the complex voltage at time instant $i$, when the voltages at instants $i-3$ through $i-1$ are known. 
	The authors in \cite{Gao2012} also test the performance of TSQPA for detecting dynamic events such as the opening of transmission lines and short-circuit faults. Robustness of TSQPA for analyzing system events for different load models has been demonstrated in \cite{Pal2015}, while it was used for conditioning and validating real PMU data in \cite{Jones2015}. However, the effectiveness of TSQPA in detecting anomalies or cyber-attacks in PMU measurements has not been investigated yet. TSQPA is emerging as a basis for real-time PMU data monitoring by some US power utilities, and therefore, it is important to evaluate its effectiveness in detecting cyber-attacks. To this end, we use TSQPA as a detector to detect anomalies due to cyber-attacks in the following way. 
	
	Applying \eqref{eq:TSQPA} on estimated voltages $\hat{x}_i$ gives the predicted voltage $\hat{x}_{(i|i-1)}$. An observation residue $r_{i,T}$ (where the subscript $T$ stands for TSQPA) at the $i^{\text{th}}$ time instant can be obtained as:
    \begin{equation}
	    r_{i,T}=\hat{x}_{(i|i-1)}-\hat{x}_{i}\label{eq:residue}
	\end{equation}
	If the magnitude of the observed residue $r_{i,T}$ exceeds a threshold, then a cyber-attack detection is declared.
	
	Finally, as a point of comparison, we also consider a higher order data-driven predictive filter, for which we similarly calculate residues to detect attacks. Details of such a filter will be given in Sec. \ref{sec:ExpSetup}.

	\subsection{Attack Design Optimization}\label{sec:AtkDesign}
	In this paper, we focus on a class of unobservable FDI attacks that aim to maximize the physical power flow on a target line subsequent to a generation re-dispatch, and possibly cause overflow \cite{Liang2015}. The attacker injects false measurements in the form of \eqref{eq:FalseMeas}, leading to false estimated states as in \eqref{eq:xbar}. Assuming the generation is known to the system operators, these false estimated states lead to false load estimations. The generation re-dispatch caused by the false loads will maximize the physical power flow on the target line. The worst-case attack can be found using an attacker-defender bi-level linear program (ADBLP) \cite{Liang2015}, wherein the first level models the attacker's objective and limitations, while the second level models the system response via DC-optimal power flow (DCOPF). The formulation of the ADBLP is given by
	\vspace{-0cm}
	\begin{subequations}\label{eq:bilevel}
		\begin{flalign}
		\hspace{-0.3cm}\underset{c, P_G^*}{\text{maximize}}\: \hspace{0.1cm} & f(P_{G}^*)\label{eq:Obj1_MaxPF}\\
		\notag \text{subject to} \\
		& A_1c\le b_1\label{eq:con_Attacker}\\
		& \left\{P_{G}^{*}\right\} =\text{arg}\left\{ \underset{P_{G}}{\text{min}}\: g(P_{G})\right\} \label{eq:OBJ_MINCOST}\\
		&\notag  \text{subject to}\\
		& \hspace{1.2cm}A_2P_G \le b_2 \label{eq:con_DCOPF}
		\end{flalign}
	\end{subequations}
	where $P_{G}$ and $P_{G}^{*}$ are vectors of generation dispatch variables and optimal generation dispatch solved by DCOPF, respectively.
	The objective function \eqref{eq:Obj1_MaxPF} maximizes the physical power flow on a target line, which is a function of generation dispatch given fixed topology and branch parameters. The attacker is constrained by \eqref{eq:con_Attacker}, including the resource limitation characterized by the $l_1$-norm of $c$, and the detection limitation characterized by the load shift caused by the attack. The system DCOPF objective \eqref{eq:OBJ_MINCOST} is to minimize the total generation cost. The DCOPF constraints are represented by \eqref{eq:con_DCOPF} that includes node balance, line limit constraints, and generation limit constraints.
	
	This ADBLP can be solved by replacing the second level problem by its Karush-Kuhn-Tucker (KKT) conditions and introducing binary variables to convert the non-convex complementary slackness conditions into mixed-integer constraints \cite{Liang2015}. The problem then becomes a single level mixed-integer linear program, and can be efficiently solved by the algorithms described in \cite{Chu2016SmartGridComm}. Alternatively, one can use a Benders' decomposition based algorithm to solve the ADBLP as introduced in \cite{Chu2019}.
	
	\section{Attack Implementation}\label{sec:AtkImplement}
	\subsection{False Measurement Creation}\label{sec:FalseMeas}
	We assume that the system performs DCOPF based on the measurements obtained at every five minutes \cite{PGEpaper}. After the system re-dispatches at time instant $i=0$, the attacker solves the ADBLP \eqref{eq:bilevel} to obtain the state attack vector $c$, and then uses $c$ to create false measurements. Although the loads at time instant $i=0$ may be different than those at the fifth minute when the system re-dispatches again, it is reasonable to assume that they will not change dramatically. Hence, the attack vector solved at $i=0$ is expected to have similar consequences to the one solved using loads at the fifth minute. Once the state attack vector $c$ is obtained, the attacker can form a measurement attack vector $d$ to create false measurements $\bar{w}$. However, it is unrealistic for the attacker to be omniscient and omnipotent. Thus, as mentioned in Sec. \ref{sec:uFDI}, we assume the attacker only controls a subset of PMUs, whose measurements are in $\mathcal{S}$. Given $c$, an attack subgraph can be constructed as in \cite{Hug2012}, consisting only of PMUs under the attacker's control. Note that here $c$ is the outcome of the ADBLP \eqref{eq:bilevel}, and hence is an attack vector on voltage angles. The measurement attack vector directly formed as $d=Hc$ will cause loads appearing at non-load buses, and possibly raise alarm at the control center. Therefore, the attacker has to solve for the final state attack vector $\tilde{c}$ that ensures the power injections at non-load buses remain unchanged, using the Newton-Raphson method as described in \cite{Liang2014}. Once $\tilde{c}$ is obtained, the measurement attack vector can be constructed as $d=H\tilde{c}$.
	
	\subsection{Attack Strategies}\label{sec:AtkStrategy}
	We consider the following two strategies for the attacker to inject false measurements:
	
	\textit{(1) Sudden attack}.  At any time instant on or before the fifth minute, the attacker injects $d$, the measurement attack vector computed at $i=0^+$, and keeps injecting $d$ afterwards. Without loss of generality, we focus on the situation where $d$ is injected at the fifth minute. Denoting $i$ as the sample number, the fifth minute is $i=9000$ assuming PMU outputs at 30 samples/sec. The false measurements in a sudden attack are given by
	\begin{equation}
	\bar{w}_i=\left\{\begin{array}{lr}
	w_i, \hspace{0.3cm} & i < 9000\\
	w_i+d, \hspace{0.3cm} & i \ge 9000
	\end{array}\right..
	\end{equation}
	A sudden attack will cause the system to re-dispatch according to the false loads, and maximize the physical power flow on the target branch. However, as we will demonstrate in Sec.~\ref{sec:Simulation}, sudden attacks can be detected by predictive filters such as TSQPA.
	
	\textit{(2) Ramping attack}. In this strategy, the attacker gradually increases the attack magnitude during the first five-minute interval, starting at $i=1$, ensuring $d$ is injected at the fifth minute, and keeps injecting $d$ afterwards. The false measurements in a ramping attack are given by
	\begin{equation}
		\bar{w}_i=\left\{\begin{array}{lr}
		w_i+\frac{i}{9000}\cdot d, \hspace{0.3cm} & i < 9000\\
		w_i+d, \hspace{0.3cm} & i \ge 9000
		\end{array}\right..
	\end{equation}
	At $t=5$ mins, the false measurements in ramping attack are identical to those in sudden attack, and hence, have the same consequences. Sec. \ref{sec:Simulation} will illustrate that predictive filters have more difficulty detecting ramping attacks due to the slow change across the 5 minute interval.

	\section{Generation of Synthetic Load Profile at PMU Time Scale}\label{sec:LoadProfile}
	To verify the proposed FDI attacks against PMU-based system operations, a realistic testbed is required; specifically, the PMU measurements used to test the BDDs must reflect realistic operating conditions. In our tests, we achieve this by simulating the dynamics of the IEEE 118 bus system with time varying loads and primary generation control. The bus-level time-series load data for this test system is generated based on a real PMU dataset that was provided by a large utility company in the southwest of the US.  The approach we adopted to create realistic load profiles is mainly based on the work described in \cite{Pinceti}. The authors present a data-driven algorithm to learn from a real dataset the spatial and temporal correlation between system loads and use the learnt model to generate new synthetic data that retains the same characteristics. In \cite{Pinceti}, the approach is demonstrated on SCADA-based, hourly load data. In this section, we detail how this technique was adapted to the learning and generation of load profiles at PMU data speeds.
	
	The utility company provided us with one week worth of PMU data for a group of neighboring substations. From the voltage and current measurements of each bus and line, we compute the loads of two substations, one at the 500kV level and one at 230kV level. Each time-series is 168 hours long, sampled at 30 samples/sec. From these two data streams we can learn the behavior of loads at different voltage levels and subsequently map them to the loads of the IEEE 118 bus system according to their voltage levels. The procedure described in the remainder of this section is followed independently for each of the two loads.
	
\begin{figure}[h]
		\centerline{\includegraphics[scale=0.5]{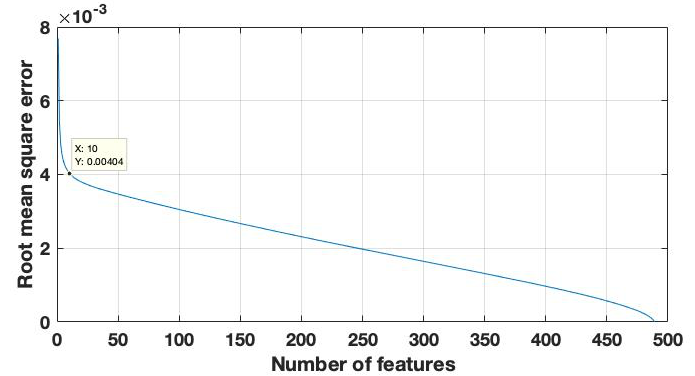}}
		\caption{Root mean squared error between $P$ and $\hat{P}$ as a function of the number of basis used.}
		\label{rmse} 
	\end{figure}

	For our simulations, we are interested in generating load data at each bus for 10 minutes. For this reason, the time-series load data for one consecutive week is broken into segments of length of 10 minutes; this results in 1008 segments, each containing 18,000 samples. The segments are then stacked to form a load matrix $P \in \mathbb{R} ^{1008\times18000}$. As shown in \cite{Pinceti}, it is possible to learn the behavior of the loads over time by factorizing the load matrix $P$ using singular value decomposition (SVD) as $P=U\Sigma V^T$. The rows of $V^T$, which are vectors of size $1\times18000$, constitute the basis of the load matrix and they correspond to archetypal \textit{temporal profiles}. 
	The synthetic loads will be generated by taking linear combinations of a subset of the first load basis (first rows of $V^T$). To determine the number of basis vectors to be used in the generative model it is useful to look at approximations of $P$, defined as $\hat{P}=U^f\times \Sigma ^f \times V^{f^T}$, where $U^f$ indicates the first $f$ columns of $U$, $\Sigma ^f$ the first $f$ columns and rows of $\Sigma$, and $V^f$ the $f$ first columns of $V$. By varying the value of $f$ (corresponding to the number of basis vectors to be used) and measuring the root mean squared error (RMSE) between $P$ and $\hat{P}$ we can determine an appropriate number of base temporal profiles to be used in the generative model. In Fig. \ref{rmse}, the error is plotted as a function of the number of basis vectors used. It can be seen that the error decreases rapidly up to $f=10$ and then it slowly reaches zero when all the basis  vectors are used. For this reason, the first 10 temporal profiles are used in the generation of the synthetic load profiles.
	
	Having identified some typical temporal load patterns, a new load profile can be created by generating a vector of coefficients and multiplying it by the set of base profiles contained in V. To compute these new coefficients we need to learn the distribution of the coefficients in the original data (e.g. the first 10 columns of $U$). Using MATLAB, it is observed that each vector of coefficients follows a different Gaussian distribution. At this point, a new matrix of load profiles for $n$ buses can be generated as: 
	\begin{equation}
		P_\text{new} = U^{10}_\text{new} \Sigma ^{10} V^{10^T}
	\end{equation}
	where $P_\text{new} \in \mathbb{R} ^{n\times18000}$, $U^{10}_\text{new} \in \mathbb{R} ^{n\times10}$ is a matrix of coefficients randomly sampled from the distributions learnt from the columns of $U$, and $\Sigma ^{10}$ and $V^{10^T}$ represent the first 10 singular values and first 10 temporal profiles obtained from the original PMU load data. To account for the spatial correlation which exists between neighboring loads, the model is modified as follows: 
	\begin{equation}\label{gen_model}
		P_\text{new} = (D U^{10}_\text{new}) \Sigma ^{10} V^{10^T}= U'^{10}_\text{new} \Sigma ^{10} V^{10^T}
	\end{equation}
	where $D \in \mathbb{R} ^{n\times n}$, and each entry $d_{i,j}$ of $D$ is given by:
	\begin{equation}\label{D}
		d_{i,j}=\begin{cases}
		1, & \text{if $i=j$}\\
		e^{-2 \text{dist}_{i,j}}, & \text{if $\text{dist}_{i,j} \leq 3$ and $i\neq j$}\\
		0, & \text{otherwise}.
		\end{cases}
	\end{equation}
	and $\text{dist}_{i,j}$ is the minimum number of branches between buses $i$ and $j$. Overall, this relation was experimentally derived in \cite{Pinceti} and was adapted to the system for which we designed the synthetic loads.

		\section{Numerical Results} \label{sec:Simulation}
	\subsection{Experiment Setup}\label{sec:ExpSetup}
	We use the IEEE 118-bus system in our simulations. The PMU placement scheme is obtained from \cite{Pal2017}. The following steps are required before we can test the performance of predictive filters for attack detection:
	\begin{enumerate}
	    \item Synthetic load profile generation: Using the model in \eqref{gen_model} on the 500kV and 230kV loads, we generate individual load profiles for 10 minutes for the loads in the IEEE 118 bus system according to their nominal voltage. Fig. \ref{load_profiles} shows the synthetic load profiles generated for two adjacent loads. As expected, they show a similar pattern over 10 minutes.
	   \begin{figure}[h]
		\centerline{\includegraphics[scale=0.7]{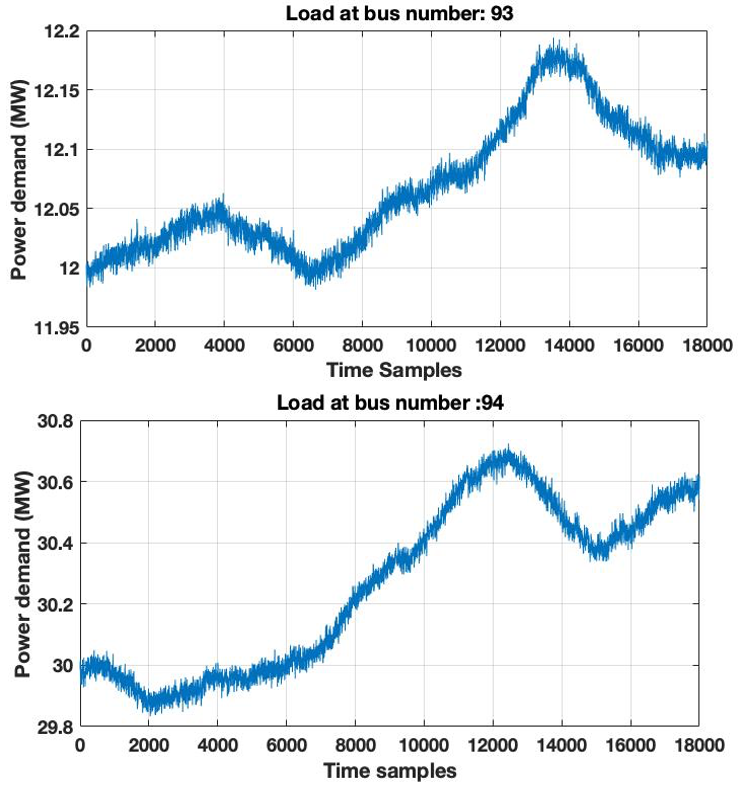}}
		\caption{Synthetic load profiles generated for two neighboring buses.}
		\label{load_profiles} 
	\end{figure}
	    \item Synthetic PMU measurements generation: Based on the synthetic loads, dynamic simulations are run in PSLF and voltage and current data are sampled 30 times per second to represent the PMU measurements. For adding noise to the synthetic PMU measurements we investigate the observation residues computed by TSQPA in the real PMU data obtained from the utility. The noise in the synthetic measurements are added in proportion to the noise in real data such that it results in similar observation residue for a no-attack scenario. The noise in magnitude and angle are selected from a Gaussian distribution of zero mean and 0.01\% standard deviation, which ensures the total vector error (TVE) to be within 1\%  \cite{PMUStandard}.
	    \item False measurements creation: A state attack vector $c$ is obtained by solving the attack design ADBLP in Sec. \ref{sec:AtkDesign} with 10\% load shift constraint. We then follow the procedure described in Sec. \ref{sec:AtkImplement} to create the measurement attack vector $d$, and subsequently the false measurements $\bar{w}$ for both sudden attack and ramping attack. The generation re-dispatch caused by the false measurements will lead to 30\% overflow on branch 54 (bus 30-38) and 22\% overflow on branch 37 (bus 8-30).
	    \item Data-driven five-sample predictive (FSP) filter: Based on the real PMU measurements that we received from the utility, we perform a moving window linear regression to learn the best coefficients of a five-sample predictive filter. This predictive filter is given by
	\begin{flalign}
	    \notag x_{(i|i-1)}= & 0.9186x_{i-1}+0.0196x_{i-2}+0.0438x_{i-3}\\
	    & +0.0058x_{i-4}+0.0122x_{i-5}. \label{eq:5_Sample_Filter}
	\end{flalign}
	\end{enumerate}
	\subsection{Attack Detection using Predictive Filters}\label{sec:AtkDetect}
\begin{figure}[ht]
	\centering
	\includegraphics[scale=0.46]{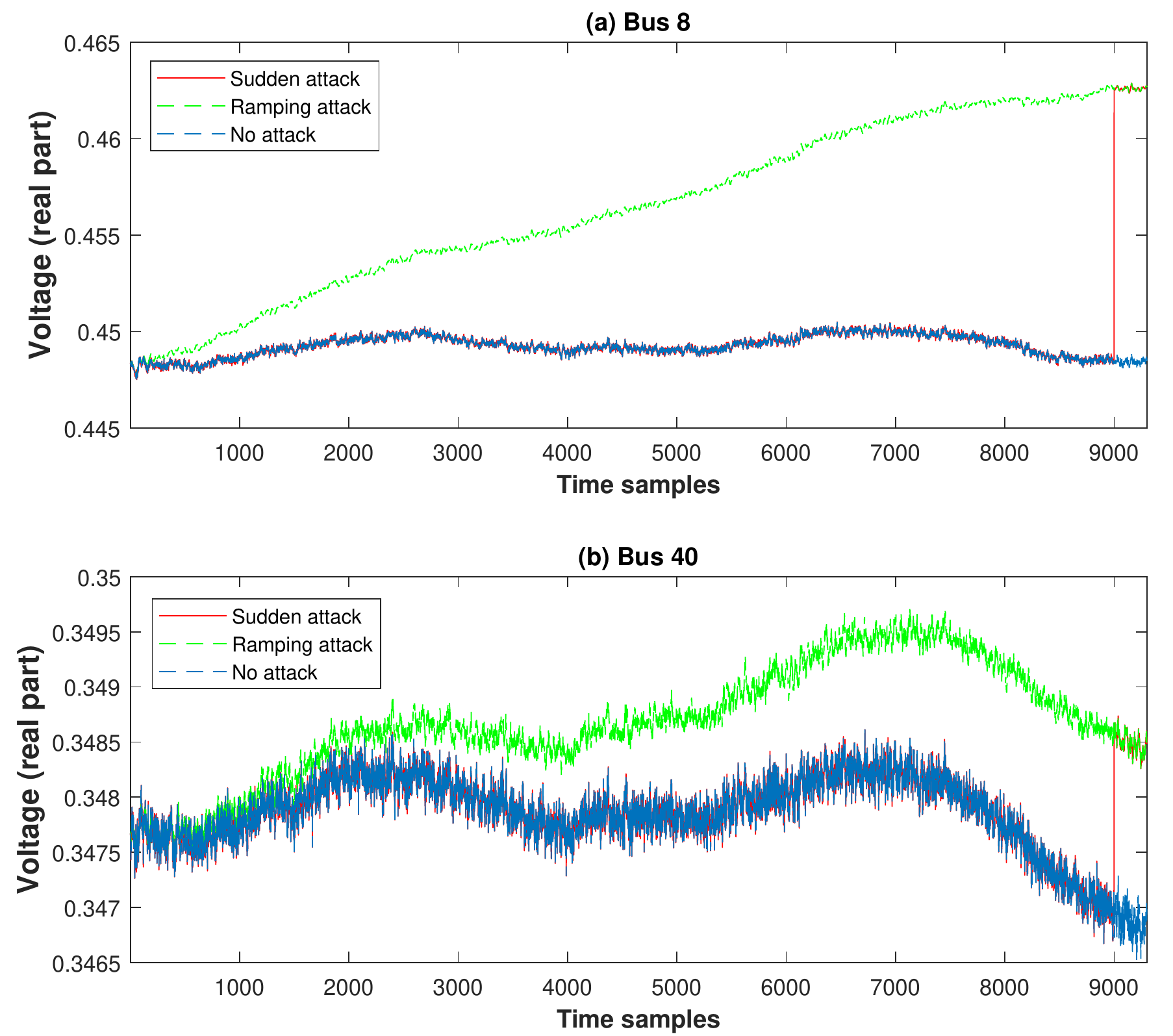} 
	\caption{Examples of false measurements at (a) bus 8; and (b) bus 40}
	\label{fig:Atk}
	\end{figure}

	We now investigate whether intelligently designed FDI attacks can be detected by predictive filters. The hypothesis of detecting an attack is that the observation residue in the presence of an attack would increase. 
	False measurements resulting from sudden and ramping attack, as well as attack-free measurements at two buses of the IEEE 118 bus system are illustrated in Fig. \ref{fig:Atk}. It can be seen that the measurements of both attack strategies are identical after 5 minutes (9,000 samples). Fig. \ref{fig:Atk}(a) shows a relatively large attack, where the attack magnitude on the real part of the voltage at bus 8 at the fifth minute is 0.0141 per unit, while Fig. \ref{fig:Atk}(b) shows a small attack at bus 40 where the attack magnitude to the real part of the voltage is merely 0.0017 per unit. 
	
\begin{figure}[ht]
	\centering
	\includegraphics[scale=0.5]{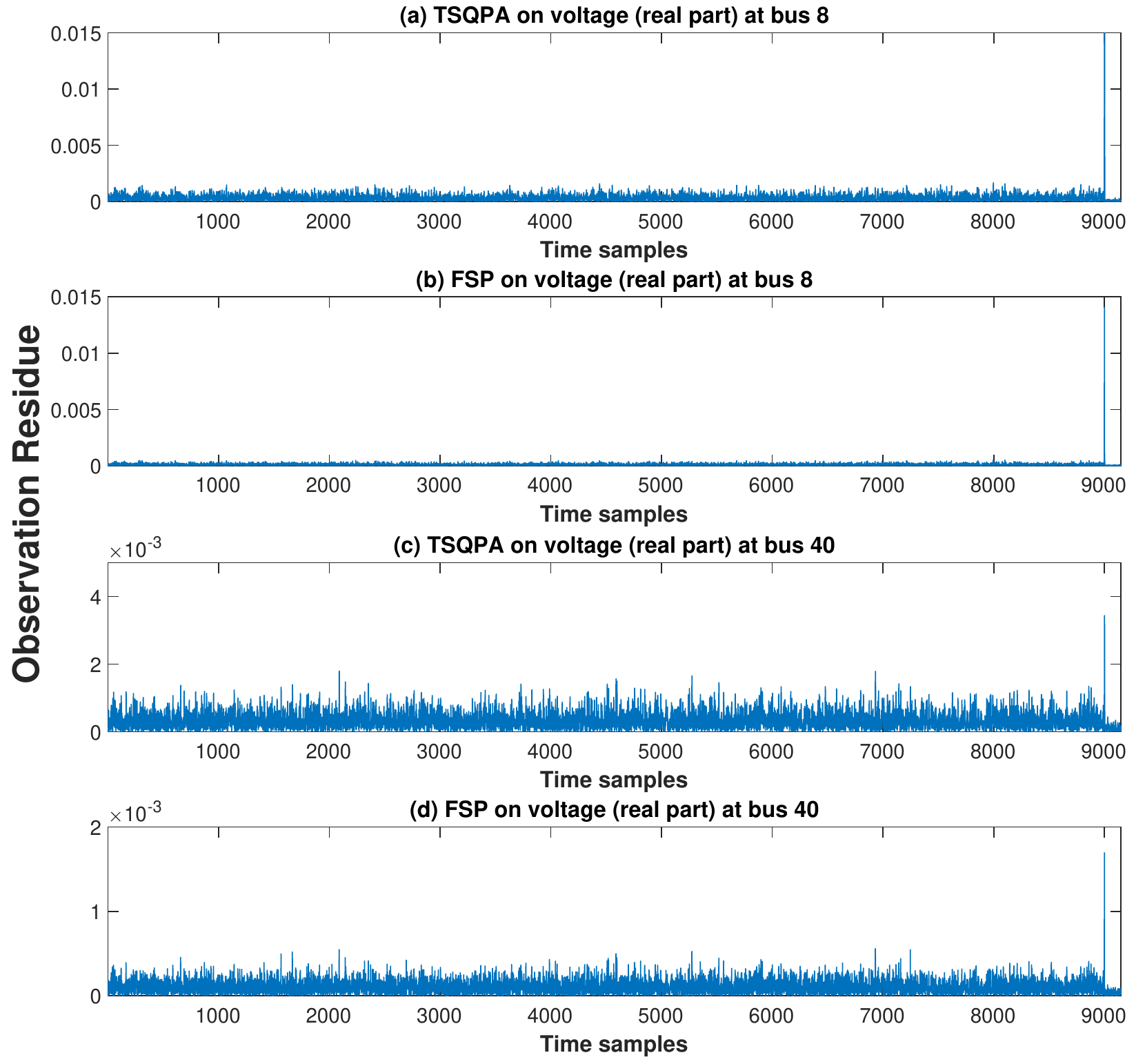}   
	\caption{Sudden attack detected by predictive filters}
	\label{fig:DetectSudden}
	\end{figure}

	Fig. \ref{fig:DetectSudden} demonstrates the observation residues when applying the  predictive filters on measurements with sudden attack. Both TSQPA and FSP give a large residue at the fifth minute when the attack is injected, indicating that they are both able to detect sudden attacks. Moreover, they can detect both the attacks at bus 8 and bus 40, even though the attack magnitude at bus 40 is much smaller. 
	
\begin{figure}[ht]
	\centering
	\includegraphics[scale=0.54]{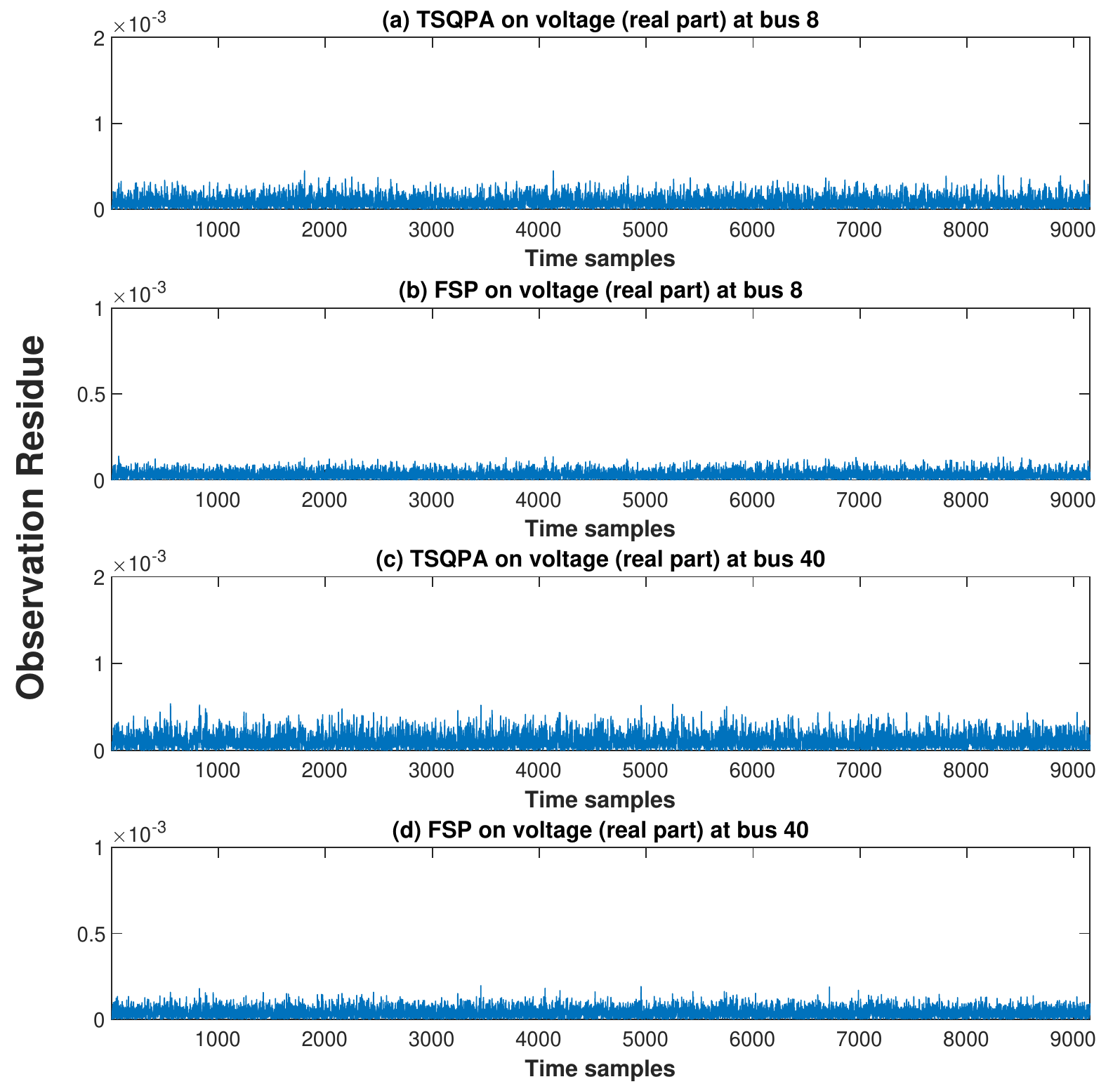}   
	\caption{Ramping attack undetected by predictive filters}
	\label{fig:DetectRamp}
	\end{figure}

	Fig. \ref{fig:DetectRamp} illustrates the observation residues obtained by applying predictive filters on measurements with ramping attack. The residues do not increase because the attack magnitude at each time instant is too small. These observations indicate that gradually ramping attacks can avoid detection by the selected predictive filters. 
	
	\section{Concluding Remarks}\label{sec:conclusion}
	In this paper, we applied two predictive filters to detect FDI attacks against PMU measurements that are unobservable by the conventional measurement residue-based bad data detector. We first created synthetic load profiles at PMU time scale that capture both temporal and spatial correlations. Using these synthetic load profiles, we then generated synthetic PMU measurements by running dynamic simulations. Subsequently, we designed test FDI attacks via a bilevel optimization approach, and created two sets of unobservable false measurements, one for sudden attack and the other for ramping attack. Finally, the false measurements are tested through a theoretically derived and a data-driven predictive filter, to see whether they can detect the attacks.
	
	The observation residues obtained from the two predictive filters for both attack strategies indicate that sudden attacks can be detected by predictive filters, while ramping attacks cannot, because the ramping attack magnitudes between time instants are smaller than those of the sudden attack. Future work will include designing detection schemes to detect ramping attacks, as well as countermeasures to mitigate FDI attacks.

	\section*{Acknowledgment}
	This material is based upon work supported by the National Science Foundation under Grant No. CNS-1449080 and the Power System Engineering Research Center (PSERC) under project S-74.
	

	%
	%

	
	
	%
	%
	%
	
	\bibliographystyle{IEEEtran}
	\bibliography{dis}

\begin{thebibliography}{10}
\providecommand{\url}[1]{#1}
\csname url@samestyle\endcsname
\providecommand{\newblock}{\relax}
\providecommand{\bibinfo}[2]{#2}
\providecommand{\BIBentrySTDinterwordspacing}{\spaceskip=0pt\relax}
\providecommand{\BIBentryALTinterwordstretchfactor}{4}
\providecommand{\BIBentryALTinterwordspacing}{\spaceskip=\fontdimen2\font plus
\BIBentryALTinterwordstretchfactor\fontdimen3\font minus
  \fontdimen4\font\relax}
\providecommand{\BIBforeignlanguage}[2]{{%
\expandafter\ifx\csname l@#1\endcsname\relax
\typeout{** WARNING: IEEEtran.bst: No hyphenation pattern has been}%
\typeout{** loaded for the language `#1'. Using the pattern for}%
\typeout{** the default language instead.}%
\else
\language=\csname l@#1\endcsname
\fi
#2}}
\providecommand{\BIBdecl}{\relax}
\BIBdecl

\bibitem{Zhao16}
J.~{Zhao}, G.~{Zhang}, K.~{Das}, G.~N. {Korres}, N.~M. {Manousakis}, A.~K.
  {Sinha}, and Z.~{He}, ``Power system real-time monitoring by using pmu-based
  robust state estimation method,'' \emph{IEEE Transactions on Smart Grid},
  vol.~7, no.~1, pp. 300--309, Jan 2016.

\bibitem{Zhang19}
Y.~{Zhang}, Y.~{Xu}, S.~{Bu}, Z.~Y. {Dong}, and R.~{Zhang}, ``Online power
  system dynamic security assessment with incomplete pmu measurements: a robust
  white-box model,'' \emph{IET Generation, Transmission Distribution}, vol.~13,
  no.~5, pp. 662--668, 2019.

\bibitem{DSA_PMU2012}
Y.~V. Makarov, P.~Du, S.~Lu, T.~B. Nguyen, X.~Guo, J.~W. Burns, J.~F.
  Gronquist, and M.~A. Pai, ``{PMU}-based wide-area security assessment:
  Concept, method, and implementation,'' \emph{IEEE Transactions on Smart
  Grid}, vol.~3, no.~3, pp. 1325--1332, Sept 2012.

\bibitem{Neyestanaki15}
M.~K. {Neyestanaki} and A.~M. {Ranjbar}, ``An adaptive pmu-based wide area
  backup protection scheme for power transmission lines,'' \emph{IEEE
  Transactions on Smart Grid}, vol.~6, no.~3, pp. 1550--1559, May 2015.

\bibitem{PMU_monitoring2010}
Y.~Zhang, P.~Markham, T.~Xia, L.~Chen, Y.~Ye, Z.~Wu, Z.~Yuan, L.~Wang, J.~Bank,
  J.~Burgett, R.~W. Conners, and Y.~Liu, ``Wide-area frequency monitoring
  network ({FNET}) architecture and applications,'' \emph{IEEE Transactions on
  Smart Grid}, vol.~1, no.~2, pp. 159--167, Sept 2010.

\bibitem{StuxNet2014}
\BIBentryALTinterwordspacing
K.~Zetter, ``An unprecedented look at {S}tuxnet, the world's first digital
  weapon,'' Nov. 2014. [Online]. Available:
  \url{https://www.wired.com/2014/11/countdown-to-zero-day-stuxnet/}
\BIBentrySTDinterwordspacing

\bibitem{UkraineAttack}
------, ``Inside the cunning, unprecedented hack of {U}kraine's power grid,''
  http://www.wired.com/2016/03/inside-cunning-unprecedented-hack-ukraines-power-grid/,
  March 2016.

\bibitem{USAttack2018_2}
\BIBentryALTinterwordspacing
B.~Ibelle, ``Russian cyberattack on {US} power grid meant to be show of power,
  researchers working to thwart the next one,'' Mar. 2018. [Online]. Available:
  \url{http://news.northeastern.edu/2018/03/21/northeastern-researchers-address-russian-power-grid-attack/}
\BIBentrySTDinterwordspacing

\bibitem{Lin_PMUcybersecurity2012}
H.~Lin, Y.~Deng, S.~Shukla, J.~Thorp, and L.~Mili, ``Cyber security impacts on
  all-{PMU} state estimator - a case study on co-simulation platform {GECO},''
  in \emph{2012 IEEE Third International Conference on Smart Grid
  Communications (SmartGridComm)}, Nov 2012, pp. 587--592.

\bibitem{Beasley2014_PMU}
C.~Beasley, G.~K. Venayagamoorthy, and R.~Brooks, ``Cyber security evaluation
  of synchrophasors in a power system,'' in \emph{2014 Clemson University Power
  Systems Conference}, March 2014, pp. 1--5.

\bibitem{Gyorgy2018}
S.~{Barreto}, M.~{Pignati}, G.~{D\'an}, J.~{Le Boudec}, and M.~{Paolone},
  ``Undetectable timing-attack on linear state-estimation by using rank-1
  approximation,'' \emph{IEEE Transactions on Smart Grid}, vol.~9, no.~4, pp.
  3530--3542, July 2018.

\bibitem{Garcia2013}
X.~{Jiang}, J.~{Zhang}, B.~J. {Harding}, J.~J. {Makela}, and A.~D.
  {Dom\'inguez-Garc\'ia}, ``Spoofing {GPS} receiver clock offset of phasor
  measurement units,'' \emph{IEEE Transactions on Power Systems}, vol.~28,
  no.~3, pp. 3253--3262, Aug 2013.

\bibitem{Gao2012}
F.~{Gao}, J.~S. {Thorp}, A.~{Pal}, and S.~{Gao}, ``Dynamic state prediction
  based on auto-regressive ({AR}) model using {PMU} data,'' in \emph{2012 IEEE
  Power and Energy Conference at Illinois}, Feb 2012, pp. 1--5.

\bibitem{Phadke2009_PMUSE}
A.~G. Phadke, J.~S. Thorp, R.~F. Nuqui, and M.~Zhou, ``Recent developments in
  state estimation with phasor measurements,'' in \emph{IEEE/PES Power Systems
  Conference and Exposition}, March 2009, pp. 1--7.

\bibitem{AburBook}
A.~Abur and A.~G. Exposito, \emph{Power System State Estimation: Theory and
  Implementation}.\hskip 1em plus 0.5em minus 0.4em\relax New York: CRC Press,
  2004.

\bibitem{Liu2009}
Y.~Liu, P.~Ning, and M.~K. Reiter, ``False data injection attacks against state
  estimation in electric power grids,'' in \emph{Proceedings of the 16th ACM
  Conference on Computer and Communications Security}, ser. CCS '09, Chicago,
  Illinois, USA, 2009, pp. 21--32.

\bibitem{Pal2015}
A.~{Pal}, ``Effect of different load models on the three-sample based quadratic
  prediction algorithm,'' in \emph{2015 IEEE Power Energy Society Innovative
  Smart Grid Technologies Conference (ISGT)}, Feb 2015, pp. 1--5.

\bibitem{Jones2015}
K.~D. {Jones}, A.~{Pal}, and J.~S. {Thorp}, ``Methodology for performing
  synchrophasor data conditioning and validation,'' \emph{IEEE Transactions on
  Power Systems}, vol.~30, no.~3, pp. 1121--1130, May 2015.

\bibitem{Liang2015}
J.~Liang, L.~Sankar, and O.~Kosut, ``Vulnerability analysis and consequences of
  false data injection attack on power system state estimation,'' \emph{IEEE
  Transactions on Power Systems}, vol.~31, no.~5, pp. 3864--3872, Sept 2016.

\bibitem{Chu2016SmartGridComm}
Z.~Chu, J.~Zhang, O.~Kosut, and L.~Sankar, ``Evaluating power system
  vulnerability to false data injection attacks via scalable optimization,'' in
  \emph{2016 IEEE International Conference on Smart Grid Communications
  (SmartGridComm)}, Nov 2016, pp. 260--265.

\bibitem{Chu2019}
\BIBentryALTinterwordspacing
------, ``Vulnerability assessment of ${N}-1$ reliable power systems to false
  data injection attacks,'' \emph{IEEE Transactions on Power Systems}, 2019,
  under review. [Online]. Available: \url{https://arxiv.org/abs/1903.07781}
\BIBentrySTDinterwordspacing

\bibitem{PGEpaper}
L.~{Zhang}, A.~{Bose}, A.~{Jampala}, V.~{Madani}, and J.~{Giri}, ``Design,
  testing, and implementation of a linear state estimator in a real power
  system,'' \emph{IEEE Transactions on Smart Grid}, vol.~8, no.~4, pp.
  1782--1789, July 2017.

\bibitem{Hug2012}
G.~Hug and J.~A. Giampapa, ``Vulnerability assessment of {AC} state estimation
  with respect to false data injection cyber-attacks,'' \emph{IEEE Transactions
  on Smart Grid}, vol.~3, no.~3, pp. 1362--1370, 2012.

\bibitem{Liang2014}
J.~Liang, O.~Kosut, and L.~Sankar, ``Cyber-attacks on {AC} state estimation:
  Unobservability and physical consequences,'' in \emph{IEEE PES General
  Meeting}, Washington, DC, July 2014.

\bibitem{Pinceti}
A.~Pinceti, O.~Kosut, and L.~Sankar, ``{Data-Driven Generation of Synthetic
  Load Datasets Preserving Spatio-Temporal Features},'' in \emph{IEEE Power and
  Energy Society General Meeting}, 2019 Accepted.

\bibitem{Pal2017}
A.~{Pal}, A.~K.~S. {Vullikanti}, and S.~S. {Ravi}, ``A {PMU} placement scheme
  considering realistic costs and modern trends in relaying,'' \emph{IEEE
  Transactions on Power Systems}, vol.~32, no.~1, pp. 552--561, Jan 2017.

\bibitem{PMUStandard}
``{IEEE} standard for synchrophasor measurements for power systems -- amendment
  1: Modification of selected performance requirements,'' \emph{{IEEE} Std
  C37.118.1a-2014 (Amendment to IEEE Std C37.118.1-2011)}, pp. 1--25, April
  2014.

\end{thebibliography}

\end{document}